\documentclass{article}
\usepackage[T1]{fontenc}
\usepackage{graphics}

\makeatletter

\newcommand{\LyX}{L\kern-.1667em\lower.25em\hbox{Y}\kern-.125emX\@}

\usepackage{a4}
\usepackage{rh}
\usepackage{vmargin}
\input{epsfig.sty}
\def\fnum@table{\tablename~{\bf\thetable}}
\def\fnum@figure{\figurename~{\bf\thefigure}}
\def\tablename{\footnotesize{\bf Table}}
\def\figurename{\footnotesize{\bf Figure}}

\makeatother

\begin{document}

\title{\textbf{\Huge Nuclear Scattering }\\
\textbf{\Huge at Very High Energies}\\
\Huge }

\author{\textbf{K. Werner}\protect\( ^{1,*}\protect \)\textbf{, H.J. Drescher\protect\( ^{1,4}\protect \),
S. Ostapchenko\protect\( ^{1,2,3}\protect \), T. Pierog\protect\( ^{1}\protect \)}\\
\\
\\
 \textit{\protect\( ^{1}\protect \)} \textit{\small SUBATECH, Université de
Nantes -- IN2P3/CNRS -- Ecole des Mines,  Nantes, France }\\
\textit{\small \protect\( ^{2}\protect \) Skobeltsyn Institute of Nuclear Physics,
Moscow State University, Moscow, Russia }\\
\textit{\small \protect\( ^{3}\protect \) Institute f. Kernphysik, Forschungszentrum
Karlsruhe, Karlsruhe, Germany}\\
\textit{\small \protect\( ^{4}\protect \) Physics Department, New York University,
New York, USA} \\
 \\
\protect\( ^{*}\protect \) Invited speaker at the International Workshop \\
on Relativistic Aspects of Nuclear Physics,\\
Caraguatatuba, Brazil, Oct. 17-20, 2000\textit{}\\
 \textit{}}
\vspace{1cm}

\maketitle
\begin{abstract}
We discuss the current understanding of nuclear scattering at very high energies.
We point out several serious inconsistencies in nowadays models, which provide
big problems for any interpretation of data at high energy nuclear collisions. 

We outline how to develop a fully self-consistent formalism, which in addition
uses all the knowledge available from studying electron-positron annihilation
and deep inelastic scattering, providing a solid basis for further developments
concerning secondary interactions.
\end{abstract}

\section{Introduction}

We will report on efforts to construct realistic and reliable models for nuclear
collisions at very high energies, say above 100 GeV cms energy per nucleon.
There are several reasons why such models are needed, related to large scale
experimental programs presently and in the future. 

In astrophysics, one of the open problems is the origin of cosmic rays at very
high energies. Whereas at low energies direct measurements are possible using
balloons or satellites, at high energies -- due to the small flux -- only indirect
measurements are possible, counting the secondary particles created in a so-called
extended air-shower. Here, theoretical models for hadron-nucleus and nucleus-nucleus
collisions play a decisive role in order to provide some information on the
origin of the primary particle. 

In nuclear physics, there are ongoing efforts to produce and to investigate
the so-called quark-gluon plasma, a new state of matter of quarks and gluons.
The SPS program at CERN has revealed many interesting results, but a final conclusion
will never be made without a profound theoretical understanding, based on realistic
microscopic models.

The outline of this paper is as follows: we will first discuss the present status
concerning models for the initial stage of a nucleus-nucleus collision. We find
that even the most advanced approaches show severe theoretical inconsistencies,
which make any interpretation of experimental data questionable. It is in fact
a well known problem which has been brought up a decade ago, but no solution
has been proposed so far. We will discuss how to solve the above-mentioned problem,
introducing a self-consistent formalism for nuclear scattering at very high
energies.

\section{Present Status}

Many popular models \cite{kai74,wer93,cap94} are based on the so-called Gribov-Regge
theory \textbf{}(GRT) \cite{gri68,gri69}. This is an effective field theory,
which allows multiple interactions to happen ``in parallel'', with the phenomenological
object called ``Pomeron'' representing an elementary interaction. Using the
general rules of field theory, one may express cross sections in terms of a
couple of parameters characterizing the Pomeron. Interference terms are crucial,
they assure the unitarity of the theory. Here one observes an inconsistency:
the fact that energy needs to be shared between many Pomerons in case of multiple
scattering is well taken into account when calculating particle production (in
particular in Monte Carlo applications), but energy conservation is not taken
care of in cross section calculations. This is a serious problem and makes the
whole approach inconsistent. 

Provided factorization works for nuclear collisions, one may employ the parton
model \cite{sjo87,wan96}, which allows to calculate inclusive cross sections
as a convolution of an elementary cross section with parton distribution functions,
with these distribution functions taken from deep inelastic scattering. In order
to get exclusive parton level cross sections, some additional assumptions are
needed, which follow quite closely the Gribov-Regge approach, encountering the
same difficulties.

Before presenting new theoretical ideas, we want to discuss the open problems
in the parton model approach and in Gribov-Regge theory somewhat more in detail.

\subsubsection*{Gribov-Regge Theory}

Gribov-Regge theory is by construction a multiple scattering theory. The elementary
interactions are realized by complex objects called ``Pomerons'', who's precise
nature is not known, and which are therefore simply parameterized: the elastic
amplitude \( T \) corresponding to a single Pomeron exchange is given as
\[
T(s,t)\sim i\, s^{\alpha _{0}+\alpha 't}\]
with a couple of parameters to be determined by experiment. Even in hadron-hadron
scattering, several of these Pomerons are exchanged in parallel. Using general
rules of field theory (cutting rules), one obtains an expression for the inelastic
cross section, 
\begin{equation}
\label{sig-part}
\sigma ^{h_{1}h_{2}}_{\mathrm{inel}}(s)=\int d^{2}b\, \left\{ 1-\exp \left( -G(s,b)\right) \right\} =\sum \sigma ^{h_{1}h_{2}}_{m}(s),
\end{equation}
with
\begin{equation}
\label{sig-m}
\sigma ^{h_{1}h_{2}}_{m}(s)=\int d^{2}b\, \frac{\left( G(s,b)\right) ^{m}}{m!}\exp \left( -G(s,b)\right) ,
\end{equation}
 where the so-called eikonal \( G(s,b) \) (proportional to the Fourier transform
of \( T(s,t) \)) represents one elementary interaction. The cross sections
\( \sigma ^{h_{1}h_{2}}_{m}(s) \) for \( m \) inelastic collisions are referred
to as topological cross sections. One can generalize to nucleus-nucleus collisions,
where corresponding formulas for cross sections may be derived.

In order to calculate exclusive particle production, one needs to know how to
share the energy between the individual elementary interactions in case of multiple
scattering. We do not want to discuss the different recipes used to do the energy
sharing (in particular in Monte Carlo applications). The point is, whatever
procedure is used, this is not taken into account in the calculation of cross
sections discussed above. So, actually, one is using two different models for
cross section calculations and for treating particle production. Taking energy
conservation into account in exactly the same way will modify the cross section
results considerably. 

This problem has first been discussed in \cite{abr92},\cite{bra90}. The authors
claim that following from the non-planar structure  of the corresponding diagrams,
conserving energy and momentum in a consistent way is crucial, and therefore
the incident energy has to be shared between the different elementary interactions,
both real and virtual ones. 

Another very unpleasant and unsatisfactory feature of most ``recipes'' for
particle production is the fact, that the second Pomeron and the subsequent
ones are treated differently than the first one, although in the above-mentioned
formula for the cross sections all Pomerons are considered to be identical.

\subsubsection*{The Parton Model}

The standard parton model approach to hadron-hadron or also nucleus-nucleus
scattering amounts to presenting the partons of projectile and target by momentum
distribution functions, \( f_{h_{1}} \) and \( f_{h_{2}} \), and calculating
inclusive cross sections for the production of parton jets with the squared
transverse momentum \( p_{\perp }^{2} \) larger than some cutoff \( Q_{0}^{2} \)
as

\[
\sigma ^{h_{1}h_{2}}_{\mathrm{incl}}=\sum _{ij}\int dp_{\perp }^{2}\int dx^{+}\int dx^{-}f^{i}_{h_{1}}(x^{+},p_{\perp }^{2})f_{h_{2}}^{j}(x^{-},p_{\perp }^{2})\frac{d\hat{\sigma }_{ij}}{dp_{\perp }^{2}}(x^{+}x^{-}s)\theta \! \left( p_{\perp }^{2}-Q^{2}_{0}\right) ,\]
where \( d\hat{\sigma }_{ij}/dp_{\perp }^{2} \) is the elementary parton-parton
cross section and \( i,j \) represent parton flavors. 

This simple factorization formula is the result of cancelations of complicated
diagrams (AGK cancelations) and hides therefore the complicated multiple scattering
structure of the reaction. The most obvious manifestation of such a structure
is the fact that at high energies (\( \sqrt{s}\gg 10 \) GeV) the inclusive
cross section in proton-(anti-)proton scattering exceeds the total one, so the
average number \( \bar{N}^{pp}_{\mathrm{int}} \) of elementary interactions
must be greater than one:
\[
\bar{N}_{\mathrm{int}}^{h_{1}h_{2}}=\sigma ^{h_{1}h_{2}}_{\mathrm{incl}}/\sigma _{\mathrm{tot}}^{h_{1}h_{2}}>1\: .\]
 The usual solution is the so-called eikonalization, which amounts to re-introducing
multiple scattering, based on the above formula for the inclusive cross section:
\begin{equation}
\label{sig-part}
\sigma ^{h_{1}h_{2}}_{\mathrm{inel}}(s)=\int d^{2}b\, \left\{ 1-\exp \left( -A(b)\, \sigma ^{h_{1}h_{2}}_{\mathrm{incl}}(s)\right) \right\} =\sum \sigma ^{h_{1}h_{2}}_{m}(s),
\end{equation}
with
\begin{equation}
\label{sig-m}
\sigma ^{h_{1}h_{2}}_{m}(s)=\int d^{2}b\, \frac{\left( A(b)\, \sigma ^{h_{1}h_{2}}_{\mathrm{incl}}(s)\right) ^{m}}{m!}\exp \left( -A(b)\, \sigma ^{h_{1}h_{2}}_{\mathrm{incl}}(s)\right) 
\end{equation}
 representing the cross section for \( n \) scatterings; \( A(b) \) being
the proton-proton overlap function (the convolution of two proton profiles).
In this way the multiple scattering is ``recovered''. This makes the approach
formally equivalent to Gribov-Regge theory, encountering therefore the same
problems: energy conservation is not at all taken care of in the above formulas
for cross section calculations.

\section{A Solution: Parton-based Gribov-Regge Theory}

\begin{figure}[htb]
{\par\centering \resizebox*{!}{0.15\textheight}{\includegraphics{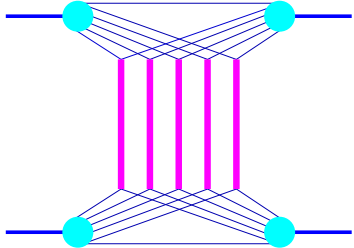}} \par}

\caption{Graphical representation of a contribution to the elastic amplitude of proton-proton
scattering. Here, energy conservation is taken into account: the energy of the
incoming protons is shared among several ``constituents'' (shown by splitting
the nucleon lines into several constituent lines), and so each Pomeron disposes
of only a fraction of the total energy, such that the total energy is conserved.\label{grtpp}}
\end{figure}

\begin{figure}[htb]
{\par\centering \resizebox*{!}{0.18\textheight}{\includegraphics{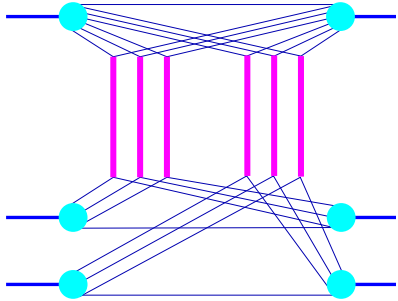}} \par}

\caption{Graphical representation of a contribution to the elastic amplitude of proton-nucleus
scattering, or more precisely a proton interacting with (for simplicity) two
target nucleons, taking into account energy conservation. Here, the energy of
the incoming proton is shared between all the constituents, which now provide
the energy for interacting with two target nucleons.\label{grtppa}}
\end{figure}

In this paper, we present a new approach for hadronic interactions and for the
initial stage of nuclear collisions, which is able to solve the above-mentioned
problems. We provide a rigorous treatment of the multiple scattering aspect,
such that questions as energy conservation are clearly determined by the rules
of field theory, both for cross section and particle production calculations.
In both (!) cases, energy is properly shared between the different interactions
happening in parallel, see fig.\ \ref{grtpp}
for proton-proton and fig.\ \ref{grtppa} for proton-nucleus collisions (generalization
to nucleus-nucleus is obvious). 
 This is the most important and new aspect of our approach, which we consider
to be a first necessary step to construct a consistent model for high energy
nuclear scattering. 

The elementary interactions, shown as the thick lines in the above figures,
are in fact a sum of a soft, a hard, and a semi-hard contribution, providing
a consistent treatment of soft and hard scattering. To some extend, our approach
provides a link between the Gribov-Regge approach and the parton model, we call
it ``Parton-based Gribov-Regge Theory''.

\section{Parton-Parton Scattering}

Let us first investigate parton-parton scattering, before constructing a multiple
scattering theory for hadronic and nuclear scattering.

We distinguish three types of elementary parton-parton scatterings, referred
to as ``soft'', ``hard'' and ``semi-hard'', which we are going to discuss
briefly in the following. The detailed derivations can be found in \cite{dre00}.

\subsubsection*{The Soft Contribution }

\begin{figure}[htb]
{\par\centering \resizebox*{!}{0.13\textheight}{\includegraphics{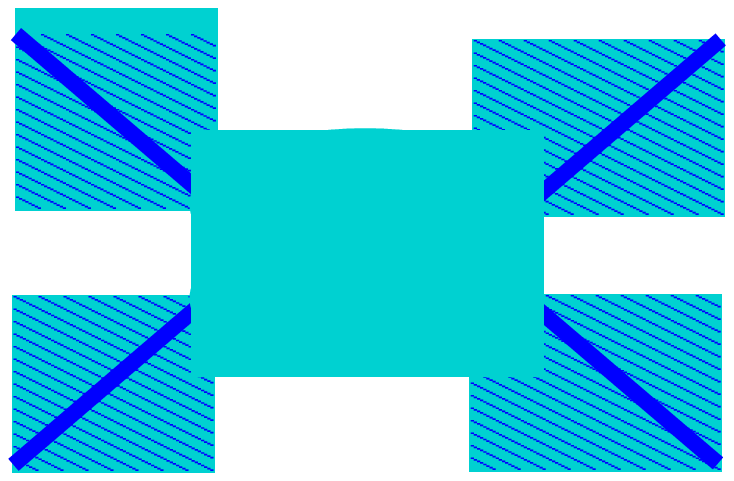}} \par}

\caption{The soft contribution.\label{fig:dsoft}}
\end{figure}
Let us first consider the pure non-perturbative contribution, where all virtual
partons appearing in the internal structure of the diagram have restricted virtualities
\( Q^{2}<Q^{2}_{0} \), where \( Q_{0}^{2}\simeq 1 \) GeV\( ^{2} \) is a reasonable
cutoff for perturbative QCD being applicable. Such soft non-perturbative dynamics
is known to dominate hadron-hadron interactions at not too high energies. Lacking
methods to calculate this contribution from first principles, it is simply parameterized
and graphically represented as a `blob', see fig. \ref{fig:dsoft}. It is traditionally
assumed to correspond to multi-peripheral production of partons (and final hadrons)
\cite{afs62} and is described by the phenomenological soft Pomeron exchange
amplitude \( T_{\mathrm{soft}}\! \left( \hat{s},t\right)  \) \cite{gri68}.
The corresponding profile function is twice the imaginary part of the Fourier
transform \( \tilde{T}_{\mathrm{soft}} \) of \( T_{\mathrm{soft}} \), divided
by the initial parton flux \( 2\hat{s} \), 
\begin{eqnarray}
D_{\mathrm{soft}}(\hat{s},b) & = & \frac{1}{2\hat{s}}2\mathrm{Im}\frac{1}{4\pi ^{2}}\int d^{2}q_{\perp }\, \exp \! \left( -i\vec{q}_{\perp }\vec{b}\right) \, T_{\mathrm{soft}}\! \left( \hat{s},-q^{2}_{\perp }\right) \nonumber \\
 & = & \frac{2\gamma _{\mathrm{part}}^{2}}{\lambda ^{\! (2)}_{\mathrm{soft}}\! (\hat{s}/s_{0})}\left( \frac{\hat{s}}{s_{0}}\right) ^{\alpha _{\mathrm{soft}}\! (0)-1}\exp \! \left( -\frac{b^{2}}{4\lambda ^{\! (2)}_{\mathrm{soft}}\! (\hat{s}/s_{0})}\right) ,\label{soft d} 
\end{eqnarray}
 with
\[
\lambda ^{\! (n)}_{\mathrm{soft}}\! (z)=nR_{\mathrm{part}}^{2}+\alpha '\! _{\mathrm{soft}}\ln \! z,\]
where \( \hat{s} \) is the usual Mandelstam variable for parton-parton scattering.
The parameters \( \alpha _{\mathrm{soft}}\! (0) \), \( \alpha '\! _{\mathrm{soft}} \)
are the intercept and the slope of the Pomeron trajectory, \( \gamma _{\mathrm{part}} \)
and \( R_{\mathrm{part}}^{2} \) are the vertex value and the slope for the
Pomeron-parton coupling, and \( s_{0}\simeq 1 \) GeV\( ^{2} \) is the characteristic
hadronic mass scale. The external legs of the diagram of fig.\ \ref{fig:dsoft}
are ``partonic constituents'', which are assumed to be quark-anti-quark pairs.

\subsubsection*{The Hard Contribution }

\begin{figure}[htb]
{\par\centering \resizebox*{!}{0.18\textheight}{\includegraphics{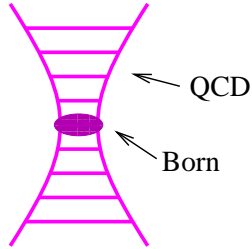}} \par}

\caption{The hard (or val-val) contribution. \label{fig:dval}}
\end{figure}

Let us now consider the other extreme, when all the processes are perturbative,
i.e. all internal intermediate partons are characterized by large virtualities
\( Q^{2}>Q^{2}_{0} \). In that case, the corresponding hard parton-parton scattering
amplitude can be calculated using the perturbative QCD techniques \cite{alt82,rey81},
and the intermediate states contributing to the absorptive part of the amplitude
can be defined in the parton basis. In the leading logarithmic approximation
of QCD, summing up terms where each (small) running QCD coupling constant \( \alpha _{s}(Q^{2}) \)
appears together with a large logarithm \( \ln (Q^{2}/\lambda ^{2}_{\mathrm{QCD}}) \)
(with \( \lambda _{QCD} \) being the infrared QCD scale), and making use of
the factorization hypothesis, one obtains the contribution of the corresponding
cut diagram for \( t=q^{2}=0 \) as the cut parton ladder cross section \( \sigma _{\mathrm{hard}}^{jk}(\hat{s},Q_{0}^{2}) \) \footnote{
Strictly speaking, one obtains the ladder representation for the process only
using axial gauge.
}, as shown in fig. \ref{fig:dval}, 
where all horizontal rungs are the final (on-shell) partons and the virtualities
of the virtual \( t \)-channel partons increase from the ends of the ladder
towards the largest momentum transfer parton-parton process (indicated symbolically
by the `blob' in the middle of the ladder):
\begin{eqnarray*}
\sigma _{\mathrm{hard}}^{jk}(\hat{s},Q_{0}^{2}) & = & \frac{1}{2\hat{s}}2\mathrm{Im}\, T^{jk}_{\mathrm{hard}}\! \! \left( \hat{s},t=0,Q_{0}^{2}\right) \\
 & = & K\, \sum _{ml}\int dx_{B}^{+}dx_{B}^{-}dp_{\bot }^{2}{d\sigma _{\mathrm{Born}}^{ml}\over dp_{\bot }^{2}}(x_{B}^{+}x_{B}^{-}\hat{s},p_{\bot }^{2})\\
 & \times  & E_{\mathrm{QCD}}^{jm}(Q_{0}^{2},M_{F}^{2},x_{B}^{+})\, E_{\mathrm{QCD}}^{kl}(Q_{0}^{2},M_{F}^{2},x_{B}^{-})\theta \! \left( M_{F}^{2}-Q^{2}_{0}\right) ,
\end{eqnarray*}
 Here \( d\sigma _{\mathrm{Born}}^{ml}/dp_{\bot }^{2} \) is the differential
\( 2\rightarrow 2 \) parton scattering cross section, \( p_{\bot }^{2} \)
is the parton transverse momentum in the hard process, \( m,l \) and \( x_{B}^{\pm } \)
are correspondingly the types and the shares of the light cone momenta of the
partons participating in the hard process, and \( M_{F}^{2} \) is the factorization
scale for the process (we use \( M_{F}^{2}=p^{2}_{\perp }/4 \)). The `evolution
function' \( E^{jm}_{\mathrm{QCD}}(Q^{2}_{0},M_{F}^{2},z) \) represents the
evolution of a parton cascade from the scale \( Q_{0}^{2} \) to \( M_{F}^{2} \),
i.e.\ it gives the number density of partons of type \( m \) with the momentum
share \( z \) at the virtuality scale \( M_{F}^{2} \), resulted from the evolution
of the initial parton \( j \), taken at the virtuality scale \( Q_{0}^{2} \).
The evolution function satisfies the usual DGLAP equation \cite{alt77} with
the initial condition \( E^{jm}_{\mathrm{QCD}}(Q^{2}_{0},Q_{0}^{2},z)=\delta ^{j}_{m}\; \delta (1-z) \).
The factor \( K\simeq 1.5 \) takes effectively into account higher order QCD
corrections.

In the following, we shall need to know the contribution of the uncut parton
ladder \( T_{\mathrm{hard}}^{jk}(\hat{s},t,Q_{0}^{2}) \) with some momentum
transfer \( q \) along the ladder (with \( t=q^{2} \)). The behavior of the
corresponding amplitudes was studied in \cite{lip86} in the leading logarithmic(\( 1/x \)
) approximation of QCD. The precise form of the corresponding amplitude is not
important for our application; we just use some of the results of \cite{lip86},
namely that one can neglect the real part of this amplitude and that it is nearly
independent on \( t \), i.e.\ that the slope of the hard interaction \( R_{\mathrm{hard}}^{2} \)
is negligible small, i.e.\ compared to the soft Pomeron slope one has \( R_{\mathrm{hard}}^{2}\simeq 0 \).
So we parameterize \( T_{\mathrm{hard}}^{jk}(\hat{s},t,Q_{0}^{2}) \) in the
region of small \( t \) as \cite{rys92}
\begin{equation}
\label{t-ladder}
T_{\mathrm{hard}}^{jk}(\hat{s},t,Q_{0}^{2})=i\hat{s}\, \sigma _{\mathrm{hard}}^{jk}(\hat{s},Q_{0}^{2})\: \exp \left( R_{\mathrm{hard}}^{2}\, t\right) 
\end{equation}

The corresponding profile function is twice the imaginary part of the Fourier
transform \( \tilde{T}_{\mathrm{hard}} \) of \( T_{\mathrm{hard}} \), divided
by the initial parton flux \( 2\hat{s} \), 
\[
D^{jk}_{\mathrm{hard}}\! \left( \hat{s},b\right) =\frac{1}{2\hat{s}}2\mathrm{Im}\tilde{T}^{jk}_{\mathrm{hard}}(\hat{s},b),\]
which gives

\begin{eqnarray}
D^{jk}_{\mathrm{hard}}\left( \hat{s},b\right) =\frac{1}{8\pi ^{2}\hat{s}}\int d^{2}q_{\perp }\, \exp \! \left( -i\vec{q}_{\perp }\vec{b}\right) \, 2\mathrm{Im}\, T_{\mathrm{hard}}^{jk}(\hat{s},-q^{2}_{\perp },Q_{0}^{2}) &  & \nonumber \\
=\sigma _{\mathrm{hard}}^{jk}\! \left( \hat{s},Q_{0}^{2}\right) \frac{1}{4\pi R_{\mathrm{hard}}^{2}}\exp \! \left( -\frac{b^{2}}{4R_{\mathrm{hard}}^{2}}\right) , &  & \label{d-val-val} 
\end{eqnarray}

In fact, the above considerations are only correct for valence quarks, as discussed
in detail in the next section. Therefore, we also talk about ``valence-valence''
contribution and we use \( D_{\mathrm{val}-\mathrm{val}} \) instead of \( D_{\mathrm{hard}} \):
\[
D^{jk}_{\mathrm{val}-\mathrm{val}}\left( \hat{s},b\right) \equiv D^{jk}_{\mathrm{hard}}\left( \hat{s},b\right) ,\]
 so these are two names for one and the same object.

\subsubsection*{The Semi-hard Contribution}

\begin{figure}[htb]
{\par\centering \resizebox*{!}{0.18\textheight}{\includegraphics{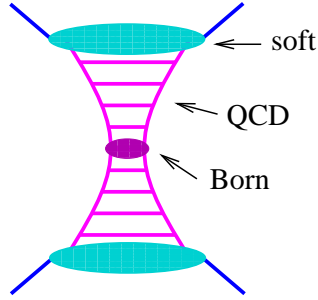}} \par}

\caption{The semi-hard ``sea-sea'' contribution: parton ladder plus ``soft ends''.\label{fig:dsemi}}
\end{figure}
The discussion of the preceding section is not valid in case of sea quarks and
gluons, since here the momentum share \( x_{1} \) of the ``first'' parton
is typically very small, leading to an object with a large mass of the order
\( Q^{2}_{0}/x_{1} \) between the parton and the proton \cite{don94}. Microscopically,
such 'slow' partons with \( x_{1}\ll 1 \) appear as a result of a long non-perturbative
parton cascade, where each individual parton branching is characterized by a
small momentum transfer squared \( Q^{2}<Q^{2}_{0} \) \cite{gri68,bak76}.
When calculating proton structure functions or high-\( p_{t} \) jet production
cross sections this non-perturbative contribution is usually included in parameterized
initial parton momentum distributions at \( Q^{2}=Q^{2}_{0} \). However, the
description of inelastic hadronic interactions requires to treat it explicitly
in order to account for secondary particles produced during such non-perturbative
parton pre-evolution, and to describe correctly energy-momentum sharing between
multiple elementary scatterings. As the underlying dynamics appears to be identical
to the one of soft parton-parton scattering considered above, we treat this
soft pre-evolution as the usual soft Pomeron emission, as discussed in detail
in \cite{dre00}.

So for sea quarks and gluons, we consider so-called semi-hard interactions between
parton constituents of initial hadrons, represented by a parton ladder with
``soft ends'', see fig. \ref{fig:dsemi}. As in the case of soft scattering,
the external legs are quark-anti-quark pairs, connected to soft Pomerons. The
outer partons of the ladder are on both sides sea quarks or gluons (therefore
the index ``sea-sea''). The central part is exactly the hard scattering considered
in the preceding section. As discussed in length in \cite{dre00}, the mathematical
expression for the corresponding amplitude is given as 
\begin{eqnarray}
iT_{\mathrm{sea}-\mathrm{sea}}(\hat{s},t) & = & \sum _{jk}\int ^{1}_{0}\! \frac{dz^{+}}{z^{+}}\frac{dz^{-}}{z^{-}}\, \mathrm{Im}\, T_{\mathrm{soft}}^{j}\! \! \left( \frac{s_{0}}{z^{+}},t\right) \, \mathrm{Im}\, T_{\mathrm{soft}}^{k}\! \! \left( \frac{s_{0}}{z^{-}},t\right) \, iT_{\mathrm{hard}}^{jk}(z^{+}z^{-}\hat{s},t,Q_{0}^{2}),\nonumber \label{t-sea-sea} 
\end{eqnarray}
with \( z^{\pm } \) being the momentum fraction of the external leg-partons
of the parton ladder relative to the momenta of the initial (constituent) partons.
The indices \( j \) and \( k \) refer to the flavor of these external ladder
partons. The amplitudes \( T_{\mathrm{soft}}^{j} \) are the soft Pomeron amplitudes
discussed earlier, but with modified couplings, since the Pomerons are now connected
to a parton ladder on one side. The arguments \( s_{0}/z^{\pm } \) are the
squared masses of the two soft Pomerons, \( z^{+}z^{-}\hat{s} \) is the squared
mass of the hard piece.

Performing as usual the Fourier transform to the impact parameter representation
and dividing by \( 2\hat{s} \), we obtain the profile function
\[
D_{\mathrm{sea}-\mathrm{sea}}\left( \hat{s},b\right) =\frac{1}{2\hat{s}}\, 2\mathrm{Im}\, \tilde{T}_{\mathrm{sea}-\mathrm{sea}}\! \left( \hat{s},b\right) ,\]
 which may be written as
\begin{eqnarray}
D_{\mathrm{sea}-\mathrm{sea}}\left( \hat{s},b\right)  & = & \sum _{jk}\int ^{1}_{0}dz^{+}dz^{-}E_{\mathrm{soft}}^{j}\left( z^{+}\right) \, E_{\mathrm{soft}}^{k}\left( z^{-}\right) \, \sigma _{\mathrm{hard}}^{jk}(z^{+}z^{-}\hat{s},Q_{0}^{2})\nonumber \\
 &  & \qquad \times \; \frac{1}{4\pi \, \lambda ^{\! (2)}_{\mathrm{soft}}(1/(z^{+}z^{-}))}\exp \! \left( -\frac{b^{2}}{4\lambda ^{\! (2)}_{\mathrm{soft}}\! \left( 1/(z^{+}z^{-})\right) }\right) \label{d-sea-sea} 
\end{eqnarray}
with the soft Pomeron slope \( \lambda ^{\! (2)}_{\mathrm{soft}} \) and the
cross section \( \sigma _{\mathrm{hard}}^{jk} \) being defined earlier. The
functions \( E_{\mathrm{soft}}^{j}\left( z^{\pm }\right)  \) representing the
``soft ends'' are defined as 
\[
\mathrm{E}^{j}_{\mathrm{soft}}(z^{\pm })=\mathrm{Im}\, T_{\mathrm{soft}}^{j}\! \! \left( \frac{s_{0}}{z^{\pm }},t=0\right) .\]
 We neglected the small hard scattering slope \( R_{\mathrm{hard}}^{2} \) compared
to the Pomeron slope \( \lambda ^{(2)}_{\mathrm{soft}} \). We call \( E_{\mathrm{soft}} \)
also the `` soft evolution'', to indicate that we consider this as simply
 a continuation of the QCD evolution, however, in a region where perturbative
 techniques do not apply any more. As discussed in \cite{dre00}, \( E_{\mathrm{soft}}^{j}\left( z\right)  \)
has the meaning of the momentum distribution of parton \( j \) in the soft
Pomeron. 

\begin{figure}[htb]
{\par\centering \resizebox*{!}{0.18\textheight}{\includegraphics{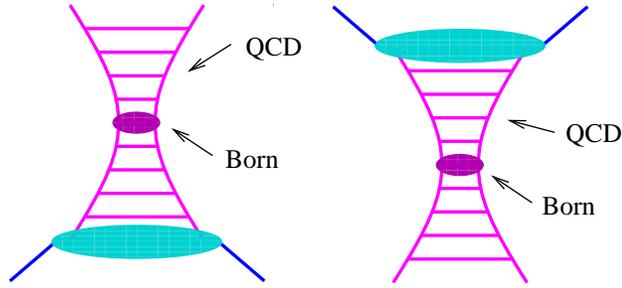}} \par}

\caption{Two ``mixed'' contributions.\label{fig:mixed}}
\end{figure}

Consistency requires to also consider the mixed semi-hard contributions with
a valence quark on one side and a non-valence participant (quark-anti-quark
pair) on the other one, see fig. \ref{fig:mixed}. 
We have
\[
iT^{j}_{\mathrm{val}-\mathrm{sea}}(\hat{s})=\int ^{1}_{0}\! \frac{dz^{-}}{z^{-}}\sum _{k}\mathrm{Im}\, T_{\mathrm{soft}}^{k}\! \! \left( \frac{s_{0}}{z^{-}},q^{2}\right) iT_{\mathrm{hard}}^{jk}\left( z^{-}\hat{s},q^{2},Q_{0}^{2}\right) \qquad \]
and
\begin{eqnarray}
D^{j}_{\mathrm{val}-\mathrm{sea}}\left( \hat{s},b\right)  & = & \sum _{k}\int ^{1}_{0}\! dz^{-}\, E_{\mathrm{soft}}^{k}\left( z^{-}\right) \, \sigma _{\mathrm{hard}}^{jk}\! \left( z^{-}\hat{s},Q_{0}^{2}\right) \label{d-val-sea} \\
 &  & \qquad \times \; \frac{1}{4\pi \, \lambda ^{\! (1)}_{\mathrm{soft}}(1/z^{-})}\exp \! \left( -\frac{b^{2}}{4\lambda ^{\! (1)}_{\mathrm{soft}}\! \left( 1/z^{-}\right) }\right) \nonumber 
\end{eqnarray}
 where \( j \) is the flavor of the valence quark at the upper end of the ladder
and \( k \) is the type of the parton on the lower ladder end. Again, we neglected
the hard scattering slope \( R^{2}_{\mathrm{hard}} \) compared to the soft
Pomeron slope. A contribution \( D^{j}_{\mathrm{sea}-\mathrm{val}}\left( \hat{s},b\right)  \),
corresponding to a valence quark participant from the target hadron, is given
by the same expression,
\[
D^{j}_{\mathrm{sea}-\mathrm{val}}\left( \hat{s},b\right) =D^{j}_{\mathrm{val}-\mathrm{sea}}\left( \hat{s},b\right) ,\]
since eq. (\ref{d-val-sea}) stays unchanged under replacement \( z^{-}\rightarrow z^{+} \)
and only depends on the total c.m.\ energy squared \( \hat{s} \) for the parton-parton
system.

\section{Hadron-Hadron Scattering}

To treat hadron-hadron scattering we use parton momentum Fock state expansion
of hadron eigenstates \cite{abr92} 
\[
|h\rangle =\sum ^{\infty }_{k=1}\frac{1}{k!}\int ^{1}_{0}\! \prod ^{k}_{l=1}\! dx_{l}\, f^{h}_{k}\! \left( x_{1},\ldots x_{k}\right) \, \delta \! \left( 1-\sum ^{k}_{j=1}\! x_{j}\right) \, a^{+}\! (x_{1})\cdots a^{+}\! (x_{k})\left| 0\right\rangle ,\]
where \( f_{k}\! \left( x_{1},\ldots x_{k}\right)  \) is the probability amplitude
for the hadron \( h \) to consist of \( k \) constituent partons with the
light cone momentum fractions \( x_{1},\ldots ,x_{k} \) and \( a^{+}\! \left( x\right)  \)
is the creation operator for a parton with the fraction \( x \). A general
scattering process is described as a superposition of a number of pair-like
scatterings between parton constituents of the projectile and target hadrons.
Then hadron-hadron scattering amplitude is obtained as a convolution of individual
parton-parton scattering amplitudes considered in the previous section and ``inclusive''
momentum distributions \( \frac{1}{n!}\tilde{F}_{h}^{(n)}\! \left( x_{1},\ldots x_{n}\right)  \)
of \( n \) ``participating'' parton constituents involved in the scattering
process(\( n\geq 1 \)), with
\[
\frac{1}{n!}\tilde{F}_{h}^{(n)}\! \left( x_{1},\ldots x_{n}\right) =\sum ^{\infty }_{k=n}\frac{1}{k!}\frac{k!}{n!\, (k-n)!}\int ^{1}_{0}\! \prod ^{k}_{l=n+1}\! \! dx_{l}\; \left| f_{k}\! \left( x_{1},\ldots x_{k}\right) \right| ^{2}\, \delta \! \left( 1-\sum ^{k}_{j=1}\! x_{j}\right) \]
 We assume that \( \tilde{F}_{h_{1}(h_{2})}^{(n)}\! \left( x_{1},\ldots x_{n}\right)  \)
can be represented in a factorized form as a product of the contributions \( F^{h}_{\mathrm{part}}(x_{l}) \),
depending on the momentum shares \( x_{l} \) of the ``participating'' or
``active'' parton constituents, and on the function \( F^{h}_{\mathrm{remn}}\! \left( 1-\sum ^{n}_{j=1}x_{j}\right)  \), 
\begin{figure}[htb]
{\par\centering \resizebox*{!}{0.1\textheight}{\includegraphics{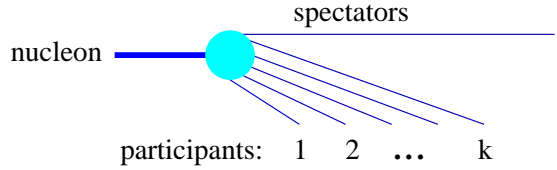}} \par}

\caption{Nucleon Fock state.\label{fig:fock}}
\end{figure}
representing the contribution of all ``spectator'' partons, sharing the remaining
share \( 1-\sum _{j}x_{j} \) of the initial light cone momentum (see fig. \ref{fig:fock}):
\begin{equation}
\label{f-part-remn}
\tilde{F}_{h}^{(n)}\! \left( x_{1},\ldots x_{n}\right) =\prod ^{n}_{l=1}\! F^{h}_{\mathrm{part}}(x_{l})\, \; F^{h}_{\mathrm{remn}}\! \left( 1-\sum ^{n}_{j=1}x_{j}\right) 
\end{equation}
 The participating parton constituents are assumed to be quark-anti-quark pairs
(not necessarily of identical flavors), such that the baryon numbers of the
projectile and of the target are conserved. Then, as discussed in detail in
\cite{dre00}, the hadron-hadron amplitude may be written as
\begin{eqnarray}
 &  & iT_{h_{1}h_{2}}(s,t)=8\pi ^{2}s\sum ^{\infty }_{n=1}\frac{1}{n!}\, \int ^{1}_{0}\! \prod ^{n}_{l=1}\! dx_{l}^{+}dx_{l}^{-}\; \prod ^{n}_{l=1}\! \left[ \frac{1}{8\pi ^{2}\hat{s}_{l}}\int \! d^{2}q_{l_{\perp }}\, iT^{h_{1}h_{2}}_{1\mathrm{I}\! \mathrm{P}}\! \left( x_{l}^{+},x_{l}^{-},s,-q_{l_{\perp }}^{2}\right) \right] \nonumber \\
 &  & F^{h_{1}}_{\mathrm{remn}}\! \! \left( 1-\sum ^{n}_{j=1}\! x_{j}^{+}\right) \, F^{h_{2}}_{\mathrm{remn}}\! \! \left( 1-\sum ^{n}_{j=1}\! x_{j}^{-}\right) \: \delta ^{(2)}\! \left( \sum ^{n}_{k=1}\! \vec{q}_{k_{\perp }}-\vec{q}_{\perp }\right) ,\label{t-hadr-hadr} 
\end{eqnarray}
where the partonic amplitudes are defined as 
\[
T_{1\mathrm{I}\! \mathrm{P}}^{h_{1}h_{2}}=T^{h_{1}h_{2}}_{\mathrm{soft}}+T^{h_{1}h_{2}}_{\mathrm{sea}-\mathrm{sea}}+T^{h_{1}h_{2}}_{\mathrm{val}-\mathrm{val}}+T^{h_{1}h_{2}}_{\mathrm{val}-\mathrm{sea}}+T^{h_{1}h_{2}}_{\mathrm{sea}-\mathrm{val}},\]
 with the individual contributions representing the ``elementary partonic interactions
plus external legs''. The soft or semi-hard sea-sea contributions are given
as
\begin{eqnarray}
T^{h_{1}h_{2}}_{\mathrm{soft}/\mathrm{sea}-\mathrm{sea}}\! \left( x^{+},x^{-},s,-q_{\bot }^{2}\right) =T_{\mathrm{soft}/\mathrm{sea}-\mathrm{sea}}\! \left( s,-q_{\bot }^{2}\right) \, F^{h_{1}}_{\mathrm{part}}(x^{+})\, F^{h_{2}}_{\mathrm{part}}(x^{-}) &  & \nonumber \\
\times \; \exp \! \left( -\left[ R_{h_{1}}^{2}+R_{h_{2}}^{2}\right] q_{\bot }^{2}\right) , &  & \label{thh} 
\end{eqnarray}
 the hard contribution is
\begin{eqnarray*}
T^{h_{1}h_{2}}_{\mathrm{val}-\mathrm{val}}\! \left( x^{+},x^{-},s,q^{2}\right)  & = & \int ^{x^{+}}_{0}\! dx_{v}^{+}\frac{x^{+}}{x_{v_{l}}^{+}}\int ^{x^{-}}_{0}\! dx_{v}^{-}\frac{x^{-}}{x_{v}^{-}}\sum _{j,k}T_{\mathrm{hard}}^{jk}\left( x_{v}^{+}x_{v}^{-}s,q^{2},Q_{0}^{2}\right) \\
 &  & \qquad \times \; \bar{F}^{h_{1},j}_{\mathrm{part}}(x_{v}^{+},x^{+}-x^{+}_{v})\, \bar{F}^{h_{2},k}_{\mathrm{part}}(x_{v}^{-},x^{-}-x_{v}^{-})\, \exp \! \left( -\left[ R_{h_{1}}^{2}+R_{h_{2}}^{2}\right] q_{l_{\perp }}^{2}\right) ,
\end{eqnarray*}
 the mixed semi-hard ``val-sea'' contribution is given as
\begin{eqnarray*}
T^{h_{1}h_{2}}_{\mathrm{val}-\mathrm{sea}}\! \left( x^{+},x^{-},s,q^{2}\right)  & = & \int ^{x^{+}}_{0}\! dx^{+}_{v}\frac{x^{+}}{x_{v}^{+}}\sum _{j}T_{\mathrm{val}-\mathrm{sea}}^{j}\left( x_{v}^{+}x^{-}s,q^{2},Q_{0}^{2}\right) \\
 &  & \qquad \times \; \bar{F}^{h_{1},j}_{\mathrm{part}}(x_{v}^{+},x^{+}-x^{+}_{v})\, F^{h_{2}}_{\mathrm{part}}(x^{-})\, \exp \! \left( -\left[ R_{h_{1}}^{2}+R_{h_{2}}^{2}\right] q_{l_{\perp }}^{2}\right) ,
\end{eqnarray*}
and the contribution ``sea-val'' is finally obtained from ``val-sea'' by
exchanging variables,
\[
T^{h_{1}h_{2}}_{\mathrm{sea}-\mathrm{val}}\! \left( x^{+},x^{-},s,q^{2}\right) =T^{h_{2}h_{1}}_{\mathrm{val}-\mathrm{sea}}\! \left( x^{-},x^{+},s,q^{2}\right) .\]
 Here, we allow formally any number of valence type interactions (based on the
fact that multiple valence type processes give negligible contribution). In
the valence contributions, we have convolutions of hard parton scattering amplitudes
\( T_{\mathrm{hard}}^{jk} \) and valence quark distributions \( \bar{F}^{j}_{\mathrm{part}} \)
over the valence quark momentum share \( x_{v}^{\pm } \) rather than a simple
product, since only the valence quarks are involved in the interactions, with
the anti-quarks staying idle (the external legs carrying momenta \( x^{+} \)
and \( x^{-} \) are always quark-anti-quark pairs). 

\begin{figure}[htb]
{\par\centering \resizebox*{!}{0.15\textheight}{\includegraphics{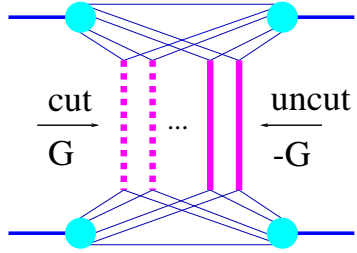}} \par}

\caption{The hadronic profile function \protect\( \gamma \protect \) expressed in terms
of partonic profile functions \protect\( G\equiv G^{h_{1}h_{2}}_{1\mathrm{I}\! \mathrm{P}}\protect \).\label{grtpabppc}}
\end{figure}

The profile function \( \gamma  \) is as usual defined as
\[
\gamma _{h_{1}h_{2}}(s,b)=\frac{1}{2s}2\mathrm{Im}\tilde{\mathrm{T}}_{h_{1}h_{2}}(s,b),\]
which may be evaluated using the AGK cutting rules with the result (assuming
imaginary amplitudes)
\begin{eqnarray}
\gamma _{h_{1}h_{2}}(s,b) & = & \sum ^{\infty }_{m=1}\frac{1}{m!}\, \int ^{1}_{0}\! \prod ^{m}_{\mu =1}\! dx_{\mu }^{+}dx_{\mu }^{-}\prod ^{m}_{\mu =1}G^{h_{1}h_{2}}_{1\mathrm{I}\! \mathrm{P}}(x_{\mu }^{+},x_{\mu }^{-},s,b)\nonumber \\
 &  & \sum ^{\infty }_{l=0}\frac{1}{l!}\, \int ^{1}_{0}\! \prod ^{l}_{\lambda =1}\! d\tilde{x}_{\lambda }^{+}d\tilde{x}_{\lambda }^{-}\prod ^{l}_{\lambda =1}-G^{h_{1}h_{2}}_{1\mathrm{I}\! \mathrm{P}}(\tilde{x}_{\lambda }^{+},\tilde{x}_{\lambda }^{-},s,b)\nonumber \\
 &  & F_{\mathrm{remn}}\left( x^{\mathrm{proj}}-\sum _{\lambda }\tilde{x}_{\lambda }^{+}\right) \, F_{\mathrm{remn}}\left( x^{\mathrm{targ}}-\sum _{\lambda }\tilde{x}_{\lambda }^{-}\right) ,\label{gam-agk-g} 
\end{eqnarray}
with \( x^{\mathrm{proj}/\mathrm{targ}}=1-\sum x^{\pm }_{\mu } \) being the
momentum fraction of the projectile/target remnant, and with a partonic profile
function \( G \) given as
\begin{eqnarray}
G^{h_{1}h_{2}}_{1\mathrm{I}\! \mathrm{P}}(x_{\lambda }^{+},x_{\lambda }^{-},s,b) & = & \frac{1}{2s}2\mathrm{Im}\, \tilde{T}^{h_{1}h_{2}}_{1\mathrm{I}\! \mathrm{P}}(x_{\lambda }^{+},x_{\lambda }^{-},s,b),\label{gss} 
\end{eqnarray}
see fig.\ \ref{grtpabppc}.
This is a very important result, allowing to express the total profile function
\( \gamma _{h_{1}h_{2}} \) via the elementary profile functions \( G^{hg_{1}h_{2}}_{1\mathrm{I}\! \mathrm{P}} \).

\section{Nucleus-Nucleus Scattering}

We generalize the discussion of the last section in order to treat nucleus-nucleus
scattering. In the Glauber-Gribov approach \cite{gla59,gri69}, the nucleus-nucleus
scattering amplitude is defined by the sum of contributions of diagrams, corresponding
to multiple scattering processes between parton constituents of projectile and
target nucleons. Nuclear form factors are supposed to be defined by the nuclear
ground state wave functions. Assuming the nucleons to be uncorrelated, one can
make the Fourier transform to obtain the amplitude in the impact parameter representation.
Then, for given impact parameter \( \vec{b}_{0} \) between the nuclei, the
only formal difference from the hadron-hadron case will be the averaging over
nuclear ground states, which amounts to an integration over transverse nucleon
coordinates \( \vec{b}^{A}_{i} \) and \( \vec{b}^{B}_{j} \) in the projectile
and in the target respectively. We write this integration symbolically as 
\begin{equation}
\int dT_{AB}:=\int \prod _{i=1}^{A}d^{2}b^{A}_{i}\, T_{A}(b^{A}_{i})\prod _{j=1}^{B}d^{2}b^{B}_{j}\, T_{B}(b^{B}_{j}),
\end{equation}
 with \( A,B \) being the nuclear mass numbers and with the so-called nuclear
thickness function \( T_{A}(b) \) being defined as the integral over the nuclear
density \( \rho _{A(B)} \): 
\begin{equation}
\label{a.11}
T_{A}(b):=\int dz\, \rho _{A}(b_{x},b_{y},z).
\end{equation}
 It is convenient to use the transverse distance \( b_{k} \) between the two
nucleons from the \( k \)-th nucleon-nucleon pair, i.e.
\[
b_{k}=\left| \vec{b}_{0}+\vec{b}^{A}_{\pi (k)}-\vec{b}^{B}_{\tau (k)}\right| ,\]
where the functions \( \pi (k) \) and \( \tau (k) \) refer to the projectile
and the target nucleons participating in the \( k^{\mathrm{th}} \) interaction
(pair \( k \)). In order to simplify the notation, we define a vector \( b \)
whose components are the overall impact parameter \( b_{0} \) as well as the
transverse distances \( b_{1},...,b_{AB} \) of the nucleon pairs,
\[
b=\{b_{0},b_{1},...,b_{AB}\}.\]
 Then the nucleus-nucleus interaction cross section can be obtained applying
the cutting procedure to elastic scattering diagram and written in the form
\begin{equation}
\label{sig-ab}
\sigma ^{AB}_{\mathrm{inel}}(s)=\int d^{2}b_{0}\int dT_{AB}\, \gamma _{AB}(s,b),
\end{equation}
where the so-called nuclear profile function \( \gamma _{AB} \) represents
an interaction for given transverse coordinates of the nucleons.

\begin{figure}[htb]
{\par\centering \resizebox*{!}{0.2\textheight}{\includegraphics{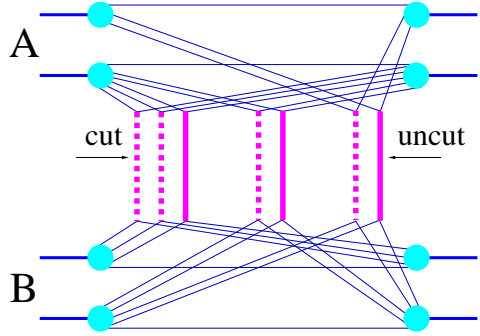}} \par}

\caption{Example for a cut multiple scattering diagram, with cut (dashed lines) and
uncut (full lines) elementary diagrams. This diagram can be translated directly
into a formula for the inelastic cross section (see text).\label{grtppaac}}
\end{figure}

The calculation of the profile function \( \gamma _{AB} \) as the sum over
all cut diagrams of the type shown in fig.\ \ref{grtppaac} does not differ
from the hadron-hadron case and follows the rules formulated
in the preceding section:

\begin{itemize}
\item For a remnant carrying the light cone momentum fraction \( x \) (\( x^{+} \)
in case of projectile, or \( x^{-} \) in case of target), one has a factor
\( F_{\mathrm{remn}}(x) \). 
\item For each cut elementary diagram (real elementary interaction = dashed vertical
line) attached to two participants with light cone momentum fractions \( x^{+} \)
and \( x^{-} \), one has a factor \( G(x^{+},x^{-},s,b) \). Apart from \( x^{+} \)
and \( x^{-} \), \( G \) is also a function of the total squared energy \( s \)
and of the relative transverse distance \( b \) between the two corresponding
nucleons (we use \( G \) as an abbreviation for \( G_{1\mathrm{I}\! \mathrm{P}}^{NN} \)
for nucleon-nucleon scattering).
\item For each uncut elementary diagram (virtual emissions = full vertical line) attached
to two participants with light cone momentum fractions \( x^{+} \) and \( x^{-} \),
one has a factor \( -G(x^{+},x^{-},s,b), \) with the same \( G \) as used
for the cut diagrams.
\item Finally one sums over all possible numbers of cut and uncut Pomerons and integrates
over the light cone momentum fractions.
\end{itemize}
So we find
\begin{eqnarray}
\gamma _{AB}(s,b) & = & \sum _{m_{1}l_{1}}\ldots \sum _{m_{AB}l_{AB}}(1-\delta _{0\Sigma m_{k}})\, \int \, \prod _{k=1}^{AB}\left\{ \prod ^{m_{k}}_{\mu =1}dx_{k,\mu }^{+}dx_{k,\mu }^{-}\, \prod ^{l_{k}}_{\lambda =1}d\tilde{x}_{k,\lambda }^{+}d\tilde{x}_{k,\lambda }^{-}\right\} \nonumber \\
 & \times  & \prod _{k=1}^{AB}\left\{ \frac{1}{m_{k}!}\frac{1}{l_{k}!}\prod _{\mu =1}^{m_{k}}G(x_{k,\mu }^{+},x_{k,\mu }^{-},s,b_{k})\prod _{\lambda =1}^{l_{k}}-G(\tilde{x}_{k,\lambda }^{+},\tilde{x}_{k,\lambda }^{-},s,b_{k})\right\} \nonumber \\
 & \times  & \prod _{i=1}^{A}F_{\mathrm{remn}}\left( x^{+}_{i}-\sum _{\pi (k)=i}\tilde{x}_{k,\lambda }^{+}\right) \prod _{j=1}^{B}F_{\mathrm{remn}}\left( x^{-}_{j}-\sum _{\tau (k)=j}\tilde{x}_{k,\lambda }^{-}\right) \label{gaminel} 
\end{eqnarray}
with 

\begin{eqnarray*}
x^{\mathrm{proj}}_{i} & = & 1-\sum _{\pi (k)=i}x_{k,\mu \, ,}^{+}\\
x^{\mathrm{targ}}_{j} & = & 1-\sum _{\tau (k)=j}x_{k,\mu }^{-}\, .
\end{eqnarray*}
 The summation indices \( m_{k} \) refer to the number of cut elementary diagrams
and \( l_{k} \) to number of uncut elementary diagrams, related to nucleon
pair \( k \). For each possible pair \( k \) (we have altogether \( AB \)
pairs), we allow any number of cut and uncut diagrams. The integration variables
\( x^{\pm }_{k,\mu } \) refer to the \( \mu ^{\mathrm{th}} \) elementary interaction
of the \( k^{\mathrm{th}} \) pair for the cut elementary diagrams, the variables
\( \tilde{x}^{\pm }_{k,\lambda } \) refer to the corresponding uncut elementary
diagrams. The arguments of the remnant functions \( F_{\mathrm{remn}} \) are
the remnant light cone momentum fractions, i.e.\ unity minus the momentum fractions
of all the corresponding elementary contributions (cut and uncut ones). We also
introduce the variables \( x^{\mathrm{proj}}_{i} \)and \( x_{j}^{\mathrm{targ}} \),
defined as unity minus the momentum fractions of all the corresponding cut contributions,
in order to integrate over the uncut ones (see below).

The expression for \( \gamma _{AB} \) sums up all possible numbers of cut Pomerons
\( m_{k} \) with one exception due to the factor \( (1-\delta _{0\Sigma m_{k}}) \):
one does not consider the case of all \( m_{k} \)'s being zero, which corresponds
to ``no interaction'' and therefore does not contribute to the inelastic cross
section. We may therefore define a quantity \( \bar{\gamma }_{AB} \), representing
``no interaction'', by taking the expression for \( \gamma _{AB} \) with
\( (1-\delta _{0\Sigma m_{k}}) \) replaced by \( (\delta _{0\Sigma m_{k}}) \):
\begin{eqnarray}
\bar{\gamma }_{AB}(s,b) & = & \sum _{l_{1}}\ldots \sum _{l_{AB}}\, \int \, \prod _{k=1}^{AB}\left\{ \prod ^{l_{k}}_{\lambda =1}d\tilde{x}_{k,\lambda }^{+}d\tilde{x}_{k,\lambda }^{-}\right\} \; \prod _{k=1}^{AB}\left\{ \frac{1}{l_{k}!}\, \prod _{\lambda =1}^{l_{k}}-G(\tilde{x}_{k,\lambda }^{+},\tilde{x}_{k,\lambda }^{-},s,b_{k})\right\} \nonumber \label{gam-bar-1} \\
 & \times  & \prod _{i=1}^{A}F^{+}\! \left( 1-\sum _{\pi (k)=i}\tilde{x}_{k,\lambda }^{+}\right) \prod _{j=1}^{B}F^{-}\left( 1-\sum _{\tau (k)=j}\tilde{x}_{k,\lambda }^{-}\right) .\label{gam-bar} 
\end{eqnarray}
One now may consider the sum of ``interaction'' and ``no interaction'',
and one obtains easily
\begin{equation}
\label{unitarity}
\gamma _{AB}(s,b)+\bar{\gamma }_{AB}(s,b)=1.
\end{equation}
Based on this important result, we consider \( \gamma _{AB} \) to be the probability
to have an interaction and correspondingly \( \bar{\gamma }_{AB} \) to be the
probability of no interaction, for fixed energy, impact parameter and nuclear
configuration, specified by the transverse distances \( b_{k} \) between nucleons,
and we refer to eq. (\ref{unitarity}) as ``unitarity relation''. But we want
to go even further and use an expansion of \( \gamma _{AB} \) in order to obtain
probability distributions for individual processes, which then serves as a basis
for the calculations of exclusive quantities.

The expansion of \( \gamma _{AB} \) in terms of cut and uncut Pomerons as given
above represents a sum of a large number of positive and negative terms, including
all kinds of interferences, which excludes any probabilistic interpretation.
We have therefore to perform summations of interference contributions -- sum
over any number of virtual elementary scatterings (uncut Pomerons) -- for given
non-interfering classes of diagrams with given numbers of real scatterings (cut
Pomerons)\cite{abr73}. Let us write the formulas explicitly. We have  
\begin{eqnarray}
\gamma _{AB}(s,b) & = & \sum _{m_{1}}\ldots \sum _{m_{AB}}(1-\delta _{0\sum m_{k}})\, \int \, \prod _{k=1}^{AB}\left\{ \prod ^{m_{k}}_{\mu =1}dx_{k,\mu }^{+}dx_{k,\mu }^{-}\right\} \nonumber \\
 & \times  & \prod _{k=1}^{AB}\left\{ \frac{1}{m_{k}!}\, \prod _{\mu =1}^{m_{k}}G(x_{k,\mu }^{+},x_{k,\mu }^{-},s,b_{k})\right\} \; \Phi _{AB}\left( x^{\mathrm{proj}},x^{\mathrm{targ}},s,b\right) ,\label{sigmanucl} 
\end{eqnarray}
 where the function \( \Phi  \) representing the sum over virtual emissions
(uncut Pomerons) is given by the following expression

\begin{eqnarray}
\Phi _{AB}\left( x^{\mathrm{proj}},x^{\mathrm{targ}},s,b\right)  & = & \sum _{l_{1}}\ldots \sum _{l_{AB}}\, \int \, \prod _{k=1}^{AB}\left\{ \prod ^{l_{k}}_{\lambda =1}d\tilde{x}_{k,\lambda }^{+}d\tilde{x}_{k,\lambda }^{-}\right\} \; \prod _{k=1}^{AB}\left\{ \frac{1}{l_{k}!}\, \prod _{\lambda =1}^{l_{k}}-G(\tilde{x}_{k,\lambda }^{+},\tilde{x}_{k,\lambda }^{-},s,b_{k})\right\} \nonumber \label{rremnant1} \\
 & \times  & \prod _{i=1}^{A}F_{\mathrm{remn}}\left( x_{i}^{\mathrm{proj}}-\sum _{\pi (k)=i}\tilde{x}_{k,\lambda }^{+}\right) \prod _{j=1}^{B}F_{\mathrm{remn}}\left( x^{\mathrm{targ}}_{j}-\sum _{\tau (k)=j}\tilde{x}_{k,\lambda }^{-}\right) .\label{rremnant} 
\end{eqnarray}
 This summation has to be carried out, before we may use the expansion of \( \gamma _{AB} \)
to obtain probability distributions. This is far from trivial, the necessary
methods are described in \cite{dre00}. To make the notation more compact, we
define matrices \( X^{+} \) and \( X^{-} \), as well as a vector \( m \),
via 
\begin{eqnarray*}
X^{+} & = & \left\{ x_{k,\mu }^{+}\right\} ,\\
X^{-} & = & \left\{ x_{k,\mu }^{-}\right\} ,\\
m & = & \{m_{k}\},
\end{eqnarray*}
which leads to
\begin{eqnarray*}
\gamma _{AB}(s,b) & = & \sum _{m}(1-\delta _{0m})\int dX^{+}dX^{-}\Omega _{AB}^{(s,b)}(m,X^{+},X^{-}),\\
\bar{\gamma }_{AB}(s,b) & = & \Omega _{AB}^{(s,b)}(0,0,0),
\end{eqnarray*}
with 
\[
\Omega _{AB}^{(s,b)}(m,X^{+},X^{-})=\prod _{k=1}^{AB}\left\{ \frac{1}{m_{k}!}\, \prod _{\mu =1}^{m_{k}}G(x_{k,\mu }^{+},x_{k,\mu }^{-},s,b_{k})\right\} \; \Phi _{AB}\left( x^{\mathrm{proj}},x^{\mathrm{targ}},s,b\right) .\]
This allows to rewrite the unitarity relation eq. (\ref{unitarity}) in the
following form,
\[
\sum _{m}\int dX^{+}dX^{-}\Omega _{AB}^{(s,b)}(m,X^{+},X^{-})=1.\]
This equation is of fundamental importance, because it allows us to interpret
\( \Omega ^{(s,b)}(m,X^{+},X^{-}) \) as probability density of having an interaction
configuration characterized by \( m \), with the light cone momentum fractions
of the Pomerons being given by \( X^{+} \) and \( X^{-} \).

\section{Virtual Emissions and Markov Chain Techniques}

What did we achieve so far? We have formulated a well defined model, introduced
by using the language of field theory, solving in this way the severe consistency
problems of the most popular current approaches. To proceed further, one needs
to solve two fundamental problems:

\begin{itemize}
\item the sum \( \Phi _{AB} \) over virtual emissions has to be performed,
\item tools have to be developed to deal with the multidimensional probability distribution \( \Omega _{AB}^{(s,b)} \),
\end{itemize}
both being very difficult tasks. Introducing new numerical techniques, we were
able to solve both problems, as discussed in very detail in \cite{dre00}.

Calculating the sum over virtual emissions (\( \Phi _{AB} \)) is done by parameterizing
the functions \( G \) as analytical functions and performing analytical calculations.
By studying the properties of \( \Phi _{AB} \), we find that at very high energies
the theory is no longer unitary without taking into account screening corrections
due to triple Pomeron interactions. In this sense, we consider our work as a
first step to construct a consistent model for high energy nuclear scattering,
but there is still work to be done.

Concerning the multidimensional probability distribution \( \Omega _{AB}^{(s,b)}(m,X^{+},X^{-}) \),
we employ methods well known in statistical physics (Markov chain techniques).
So finally, we are able to calculate the probability distribution \( \Omega _{AB}^{(s,b)}(m,X^{+},X^{-}) \),
and are able to generate (in a Monte Carlo fashion) configurations \( (m,X^{+},X^{-}) \)
according to this probability distribution.

\section{Summary}

What are finally the principal features of our basic results, summarized in
eqs. (\ref{sig-ab}, \ref{sigmanucl}, \ref{rremnant})? Contrary to the traditional
treatment (Gribov-Regge approach or parton model), all individual elementary
contributions \( G \) depend explicitly on the light-cone momenta of the elementary
interactions, with the total energy-momentum being precisely conserved. Another
very important feature is the explicit dependence of the screening contribution
\( \Phi _{AB} \) (the contribution of virtual emissions) on the remnant momenta.
The direct consequence of properly taking into account energy-momentum conservation
in the multiple scattering process is the validity of the so-called AGK-cancelations
in hadron-hadron and nucleus-nucleus collisions in the entire kinematical region. 

The formulas (\ref{sig-ab}, \ref{sigmanucl}, \ref{rremnant}) allow to develop
a consistent scheme to simulate high energy nucleus-nucleus interactions. The
corresponding Monte Carlo procedure is exactly based on the cross section formulas
so that the entire model is fully self-consistent. 

\bibliographystyle{pr2}
\bibliography{a}

\end{document}